\documentclass[aps,prl,twocolumn,showpacs,preprintnumbers,nofootinbib,amsmath,amssymb]{revtex4}

\usepackage{graphicx}
\usepackage{dcolumn}
\usepackage{bm}
\usepackage{amsmath}

\newcommand{\beq}{\begin{eqnarray}}
\newcommand{\eeq}{\end{eqnarray}}

\newcommand{\real}{{\sf I}\kern-.12em{\sf R}}
\newcommand{\comp}{{\sf I}\kern-.50em{\sf C}}
\newcommand{\unity}{{\sf I}\kern-.54em{\sf 1}}

\newcommand{\tef}{\theta_{\rm eff}}
\newcommand{\tief}{\theta_{I \rm eff}}
\newcommand{\esb}{\vec E \cdot \vec B}
\newcommand{\eisb}{\vec E_I \cdot \vec B}
\newcommand{\ceb}{\chi_{CP}\,}
\newcommand{\avq}{\langle Q \rangle}
\newcommand{\avqs}{\langle Q^2 \rangle}

\def\spose#1{\hbox to 0pt{#1\hss}}
\def\ltapprox{\mathrel{\spose{\lower 3pt\hbox{$\mathchar"218$}}
 \raise 2.0pt\hbox{$\mathchar"13C$}}}

\begin{document}

\title{Susceptibility of the QCD vacuum to 
CP-odd electromagnetic background fields}
\author{Massimo D'Elia and Marco Mariti}
\affiliation{
Dipartimento di Fisica dell'Universit\`a
di Pisa and INFN - Sezione di Pisa,\\ Largo Pontecorvo 3, I-56127 Pisa, Italy}
\email{delia@df.unipi.it,mariti@df.unipi.it}
\author{Francesco Negro}
\affiliation{Dipartimento di Fisica dell'Universit\`a
di Genova and INFN - Sezione di Genova,\\
 Via Dodecaneso 33, I-16146 Genova, Italy}
\email{fnegro@ge.infn.it}
\date{\today}

\begin{abstract}
We investigate two flavor QCD 
in presence of CP-odd electromagnetic background fields and
determine, by means of lattice QCD simulations, the induced 
effective $\theta$ term  
to the first order in $\vec E \cdot \vec B$. We employ 
a rooted staggered discretization and study lattice spacings down to 
$0.1$ fm and Goldstone pion masses around 480 MeV.
In order to deal with a positive
measure, we consider purely imaginary electric
fields and real magnetic fields, then exploiting analytic continuation.
Our results are relevant to a description of the effective pseudoscalar 
QED-QCD interactions.
\end{abstract}

\maketitle

Quantum ChromoDynamics (QCD) may contain
interactions which violate CP, the symmetry under charge conjugation and parity, 
corresponding to a term 
$- i\, \theta\, Q$ in the Euclidean action, where
\beq
Q = \int d^4 x\, q(x) = 
\int d^4 x \frac{g^2}{64\pi^2} G_{\mu\nu}^a(x)
\tilde G_{\mu\nu}^a(x)
\label{thetaterm}
\eeq
is the topological charge operator, 
$G_{\mu\nu}^a$ is the non-Abelian gauge field strength and
$\tilde G_{\mu\nu}^a = 
\epsilon_{\mu\nu\rho\sigma} G_{\rho\sigma}^a$.
Experimental upper bounds on $\theta$ 
are quite stringent,
$|\theta| \lesssim 10^{-10}$~\cite{thetabound1,thetabound2}. 
Nevertheless, $\theta$ related effects
play an important role in strong interactions
and are generally linked to fluctuations of $Q$, which affect,
through the axial anomaly, the balance of chirality.

A significant interest is related to the possibility
that local effective variations of $\theta$, corresponding to
topological charge fluctuations, may induce detectable phenomena
in presence of magnetic fields as strong as
those produced in the early phases of non-central
heavy ion collisions, reaching up
to $10^{15}$ Tesla at LHC. 
According to the so-called chiral magnetic 
effect (CME)~\cite{cme0,cme1,cme2}, 
the net unbalance of chirality
induced by the topological background would lead, in presence 
of a magnetic field strong enough to align quark spins,
to a net separation of electric charge along the field direction.

The physics of strong interactions in presence
of electromagnetic (e.m.) backgrounds has attracted
much interest also in relation to the possible effects 
on the QCD vacuum and on the QCD phase diagram,
stimulating many 
model computations and
lattice simulations.
Such effects may be relevant in various contexts: 
magnetic fields as large as
of the order of 
$10^{16}$ Tesla may have been 
produced at the cosmological electroweak phase 
transition~\cite{Vachaspati:1991nm};
large magnetic fields
are also expected in compact
astrophysical objects such as  
magnetars~\cite{magnetars}.

One aspect, emerging from lattice simulations
with dynamical fermions~\cite{lat1,lat2,lat3,lat4,lat5,latrev}
and from model studies~\cite{shov,anisotropic,galilo},
is
that e.m. background fields, even if directly coupled only to 
charged particles, may have a significant influence, via quark loop effects,
also on gluon fields.
In this study, in an attempt to better clarify such issue,
we will investigate how the explicit breaking of some symmetry
by the e.m. background field propagates 
to gluon fields, considering the particular case of CP symmetry.

Let us consider QCD in presence of a constant and uniform
e.m. field such that 
$F_{\mu\nu} \tilde F_{\mu\nu} \propto \esb \neq 0$:
such background is expected to induce 
an effective CP-violating interaction in the gluon sector, 
$\tef \frac{g^2}{64\pi^2} G_{\mu\nu}^a(x) \tilde G_{\mu\nu}^a(x)$. 
$\tef$ must be an odd function of $\esb$, 
hence  
we can write
\beq
\tef \simeq \ceb e^2 \esb + O( (\esb)^3 )
\label{deftef}
\eeq
 where
$\ceb$ is a sort of susceptibility 
of the QCD vacuum to CP-breaking e.m. fields.

The effect that we want to study is in some sense complementary to the CME, 
where a CP-violating non-Abelian background
leads to charge separation, hence to an electric
field, parallel to a background magnetic field:
in that case CP violation propagates from the gluonic to the e.m.
sector, i.e. opposite to what we investigate here. 
In fact, $\ceb$ is directly related 
of the 
effective pseudoscalar QED-QCD interaction,
$\ceb q(x) e^2 \esb$~\cite{musak,elze1,elze2,mueller}; 
in particular, to connect with
the notation of Ref.~\cite{mueller}, we have $\ceb = \kappa/2$, with
$\kappa$ defined as in Eq.~(5) of Ref.~\cite{mueller}.

The purpose
of this study is to furnish a first determination
of $\ceb$ based on lattice QCD simulations. To that aim we 
simulate
QCD in presence of uniform e.m. background fields such
that $\esb \neq 0$, determining the 
induced $\tef$ by studying the topological charge distribution.

{\it The method } --
Electromagnetic fields
enter the QCD lagrangian by modifying 
the covariant derivative
of quarks,
$D_\mu = 
\partial_\mu + i\, g A^a_\mu T^a + i\, q A_\mu$, where 
$A_\mu$ is the e.m. gauge potential and $q$ 
is the quark electric charge. 
That can be implemented by adding 
proper $U(1)$ phases $u_\mu(n)$ to the $SU(3)$ parallel transports,
$U_\mu(n) \to u_\mu(n) U_\mu(n)$, where $n$ is a lattice site.

A constant and uniform e.m. field, with a single non vanishing 
component
$F_{\mu\nu} = F$,
can be realized by a 
potential
$
A_\nu = F  x_\mu$
and 
$A_\rho = 0\ {\rm for}\ \rho\neq\nu
$.
In presence of periodic boundary conditions (b.c.), 
as usual in lattice simulations,
$F$ must be integer multiple of a minimum 
quantum 
\beq
f 
= {2 \pi}/({ q a^{2} L_\mu L_\nu})\, , 
\label{fquant}
\eeq
and
proper b.c. must 
be chosen for fermions,
to preserve gauge invariance~\cite{wiese}. The corresponding  
$U(1)$ links are
\beq
u_\nu^{(q)}(n) = e^{i\, a^2 q F\, n_\mu}
\ \, ;\ \ \  
u_\mu^{(q)}(n)|_{n_\mu = L_\mu} = e^{-i\, a^2 q L_\mu F\, n_\nu} \, 
\label{u1field}
\eeq
and $u_\rho(n) = 1$ otherwise, $L_\mu$ being the number of lattice sites in the
$\mu$ direction. 
In presence of various
non-vanishing components of $F_{\mu\nu}$,
the definition above generalizes by summing, for
each $U(1)$ phase, the contributions from 
the various non-vanishing components of $F_{\mu\nu}$.

We have considered two flavor QCD with fermions discretized
in the standard rooted staggered formulation. In the corresponding functional integral,
each quark is described
by the fourth root of the fermion matrix determinant:
\beq
Z \equiv \int \mathcal{D}U e^{-S_{G}} 
\det M^{1\over 4} [U,q_u]
\det M^{1\over 4} [U,q_d]
\:
\label{partfun1}
\eeq
\begin{eqnarray}
M_{i,j} = a m \delta_{i,j} 
&+& {1 \over 2} \sum_{\nu=1}^{4} \eta_\nu(i) \left(
u_\nu^{(q)}(i)\ U_{\nu}(i) \delta_{i,j-\hat\nu}
\right. \nonumber \\ &-& \left.
u^{*(q)}_\nu{(i - \hat\nu)}\ U^{\dag}_\nu{(i-\hat\nu)} \delta_{i,j+\hat\nu} 
\right) \:.
\label{fmatrix1}
\end{eqnarray}
$\mathcal{D}U$ is the functional integration over the non-Abelian gauge link
variables, $S_G$ is the plaquette action,  
$i$ and $j$ refer to lattice sites and
$\eta_{\nu}(i)$ are the staggered
phases. The choice for the quark charges is standard,
$q_u = 2 |e|/3$ and $q_d = - |e|/3$, leading to a quantization
in units of $f = {6 \pi}/({ |e| a^{2} L_\mu L_\nu})$ for each 
component of $F_{\mu\nu}$.

The addition of $U(1)$ phases to the standard $SU(3)$ link variables leaves
the spectrum of $M\hspace{-2pt} -\hspace{-2pt} m\, {\rm Id}$ purely imaginary and symmetric under conjugation: that guarantees
$\det M > 0$, hence the feasibility of numerical simulations. However, it is easy to realize
that $F_{0i} \neq 0$, as defined above, corresponds
to a purely imaginary electric field in Minkowski space~\cite{shintani,alexandru}.

In order to have a real electric field, 
one should introduce imaginary components for the gauge potential in Euclidean space,
but that would take the $u_\mu$ variables out of the $U(1)$ group and make
the quark determinant complex: such sign problem would 
hinder numerical
simulations. This is expected in view of the phenomenon that
we want to explore, since also the formulation of QCD at
non-vanishing $\theta$ suffers from a sign problem.

A possible strategy, adopted in lattice studies of the electric polarizabilities of hadrons,
is to keep the electric field purely imaginary, to avoid 
the sign problem,
then exploiting analytic continuation. Following such approach, 
we adopt the
definition in Eq.~(\ref{u1field}) 
for all components of $F_{\mu\nu}$, corresponding 
to real magnetic fields $\vec B$ and imaginary electric fields $\vec E  = i\, \vec E_I$. 
As a consequence, we expect a purely imaginary effective
parameter $\tef = i\, \tief$. 
On the other hand, working with an imaginary $\theta$ 
is also one of the possible approaches to the study of $\theta$-dependence in QCD~\cite{azcoiti, alles_1, aoki_1, vicari_im, tctheta}. 

The presence of an imaginary $\theta_I$ adds a factor
$\exp( \theta_I Q)$ to the path integral measure, 
shifting 
distribution of $Q$ by an amount which,
at the 
linear order in $\theta_I$, is governed by the topological 
susceptibility $\chi$ at $\theta_I = 0 $:
\beq
\avq_{\theta_I} \simeq V\, \chi\, \theta_I = \avqs_{\theta = 0}\, 
\theta_I \, , 
\eeq  
$V$ being the spacetime volume. 
That suggests to determine 
the effective $\theta_I$ produced by a given e.m. field 
as
\beq
\tief \simeq\,
{\avq (\vec E_I, \vec B)}/{\avqs_0}\,  +\,  O( (\eisb)^3 )
\label{deftief}
\eeq
where by $\langle \cdot \rangle_0$ we mean the average taken 
at zero e.m. field.
Corrections to Eq.~(\ref{deftief}) will be negligible
at least in the region of small $\tief$ which 
is relevant to Eq.~(\ref{deftef}).
In the following we will
show that the so defined $\tief$ is indeed an odd function of 
$\eisb$ alone and, assuming analyticity
for small enough background fields,
we shall determine the susceptibility $\ceb$ defined in Eq.~(\ref{deftef})
from the small field behavior of $\tief$.

\begin{table}[t!]
\begin{center}
\begin{tabular}{|c|c|c|c|c|}
\hline
lattice size & $a$ (fm) & $\beta$ & $a m$ & $\ceb$(GeV$^{-4}$) \\ 
\hline
\hline
L = 16\ & 0.094\ &  5.6286\ & 0.008318\ & 5.62(55)\ \\
\hline
L = 16\ & 0.113\ &  5.5829\ & 0.01048\ & 5.39(39)\ \\
\hline
L = 12, 24\ & 0.141\  & 5.527\ & 0.0146\ & 5.54(47)\ \\
\hline
L = 12, 18\ & 0.188\ &  5.453\ & 0.02627\ & 2.89(39)\ \\
\hline
L = 12, 16\ & 0.282\  &  5.35\ & 0.075\ & 1.05(12)\ \\
\hline
\end{tabular}
\end{center}
\caption{Lattice parameters and results for $\ceb$.}
\label{tab1}
\end{table}

\begin{figure}[t!]
\includegraphics*[width=0.49\textwidth]{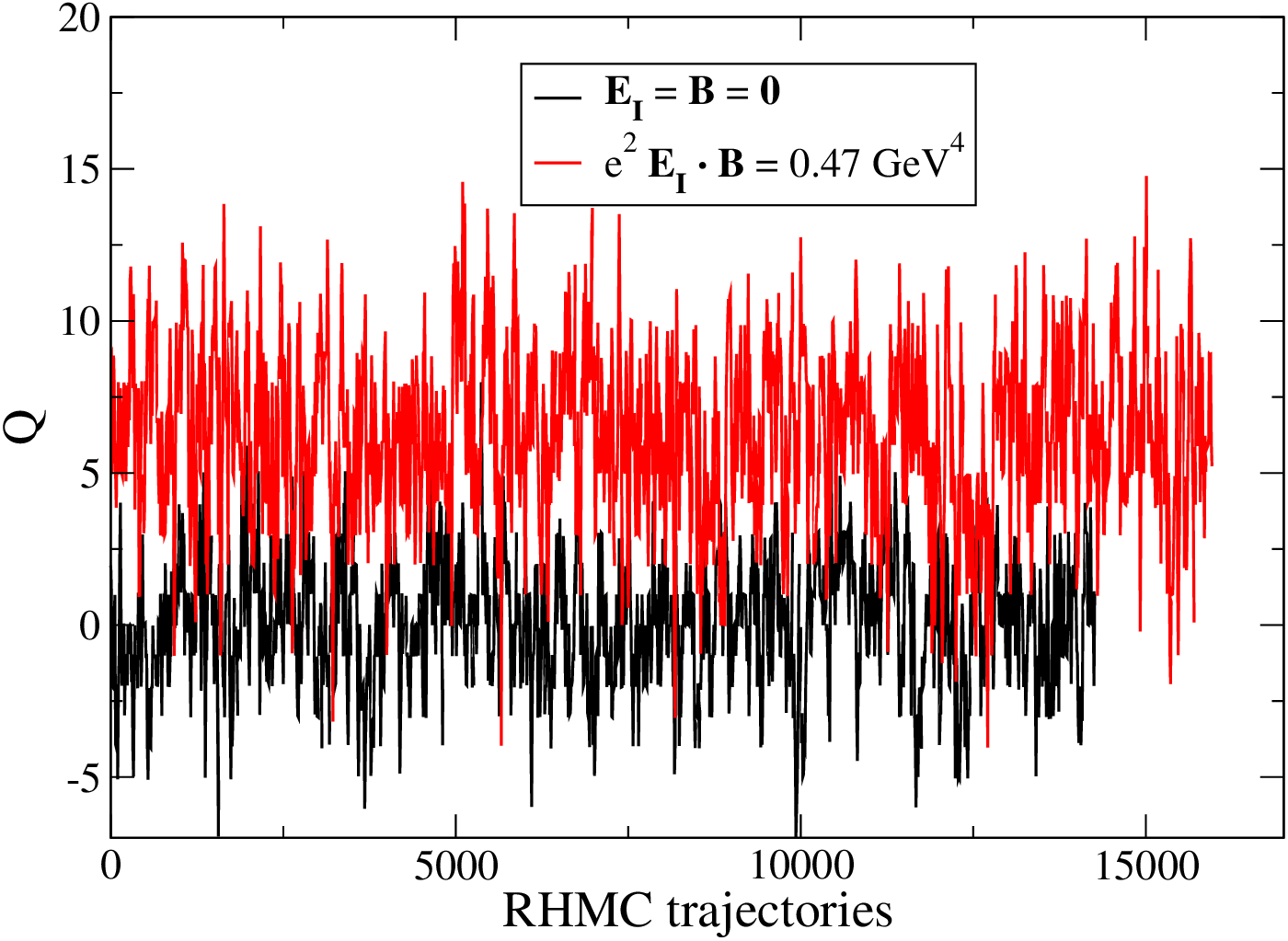}
\caption{Monte-Carlo history of $Q$, measured after 30 cooling steps,
for two different e.m. backgrounds, on a $16^4$ lattice;
$m_\pi \simeq 480$ MeV and $a \simeq 0.113$ fm.
}
\label{fig:1b}
\end{figure}

{\it Results} -- 
We have studied
$N_f = 2$ QCD at $T = 0$
for a fixed pseudo-Goldstone pion mass
$m_\pi \simeq 480$ MeV~\cite{taste}. We have explored different
lattice spacings, by tuning the inverse gauge coupling
$\beta$ and $a m$ according to what reported
in Ref.~\cite{blum}, and different 
symmetric lattice sizes (see Table~\ref{tab1}).
The lattice spacing is not modified by the presence 
of a purely magnetic background~\cite{lat3}; we have verified, 
by measuring the static quark-antiquark potential,
that this is the case also when both
$\vec E_I$ and $\vec B$ are non-zero~\cite{footnote1}.
We have adopted a 
Rational Hybrid Monte-Carlo (RHMC)
algorithm implemented on GPU cards~\cite{gpu}, with statistics 
of $O(10{\rm K})$ molecular dynamics time units
for each run. 

\begin{figure}[t!]
\includegraphics*[width=0.49\textwidth]{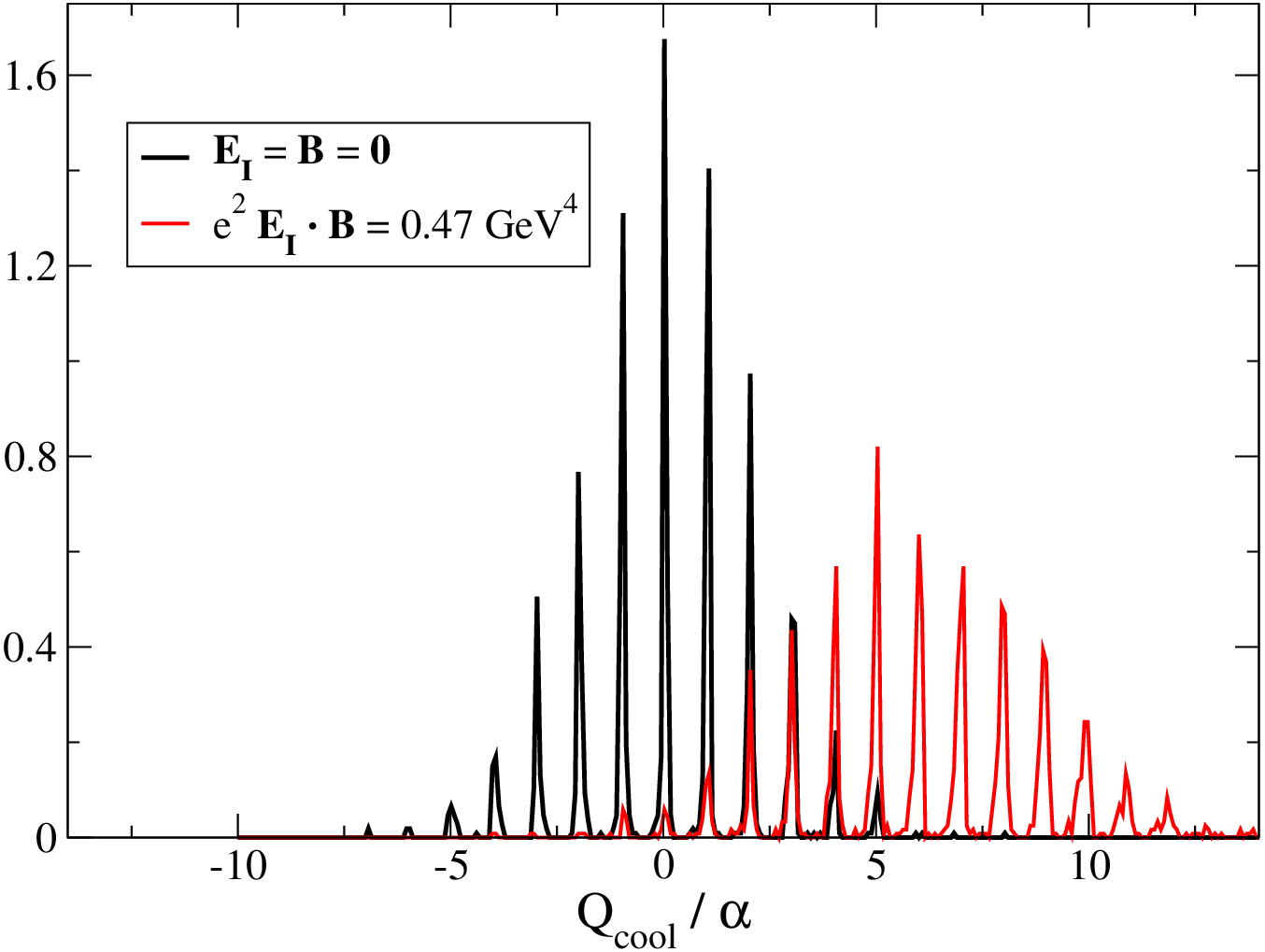}
\caption{Distribution of the topological charge shown in 
Fig.~\ref{fig:1b}, $\alpha = 0.947$ (see text).
} 
\label{fig:1}
\end{figure}

The total number of about 300~K topological charge measurements
compelled us to adopt
a relatively cheap determination
of $Q$, based on a gluonic definition
\beq
q_L(x) = {{-1} \over {2^9 \pi^2}} 
\sum_{\mu\nu\rho\sigma = \pm 1}^{\pm 4} 
{\tilde{\epsilon}}_{\mu\nu\rho\sigma} \hbox{Tr} \left( 
\Pi_{\mu\nu}(x) \Pi_{\rho\sigma}(x) \right) \; ,
\label{eq:qlattice}
\eeq
measured after cooling~\cite{cooling,vicari_rep}, 
i.e.~minimization of the gauge action to eliminate ultraviolet (UV) artifacts; 
${\tilde{\epsilon}}_{\mu\nu\rho\sigma} = {{\epsilon}}_{\mu\nu\rho\sigma}
$ for positive directions  and ${\tilde{\epsilon}}_{\mu\nu\rho\sigma} =
- {\tilde{\epsilon}}_{(-\mu)\nu\rho\sigma}$. 
After cooling, the charge is divided by a constant factor
$\alpha$, typically very close to one, so that its distribution
gets peaked around integer values (see e.g. Fig.~\ref{fig:1}), then 
$Q$ is fixed to the closest integer.

For each run we have verified the stability of results against the number
of cooling sweeps $n_{\rm cool}$, then fixing $n_{\rm cool}$ inside 
a well defined plateau (e.g. $n_{\rm cool} = 30$
for the finest and 60 for the coarsest spacing).
Cooling is known to provide results comparable with 
improved fermionic definitions as the continuum limit is 
approached~\cite{vicari_rep,zhang_ov,teper_ov,bruck}, however, due to the unusual
conditions adopted in our investigation, we have checked this fact
on subsets of our configurations.
For $a = 0.094$ fm and $\eisb \simeq 0.22$ GeV$^4$ we obtain,
on a set of 30 decorrelated configurations,
$\langle Q_{ov} \rangle / \langle Q_{30} \rangle \simeq 0.81(8)$,
where
$Q_{ov}$ counts the difference between left and 
right-handed zero modes of the 
Dirac overlap operator
and 
$Q_{n}$ refers to $n$ cooling sweeps; 
on the full set of about 8000 configurations we have instead
$\langle Q_{60} \rangle / \langle Q_{15} \rangle \simeq 0.987(9)$.
The agreement
is reasonable since, at this lattice spacing,
small topological objects may not correspond to exact zero 
modes~\cite{teper_ov}.
At the same $a$, 
autocorrelation times for $Q$, which grow critically 
when approaching the continuum limit, are around
100 RHMC trajectories.

As a first illustration of our results, 
in Fig.~\ref{fig:1b} we show 
the Monte-Carlo history of $Q$ for two numerical simulations, performed respectively
at zero and non-zero e.m. field,
for $a \simeq 0.113$ fm. 
The non-zero
e.m. field corresponds to $\vec E_I = \vec B = 3\, f\, \hat z$, with
$f$ defined in Eq.~(\ref{fquant}),  
i.e.~$e^2 \eisb \simeq 0.47\, {\rm GeV}^4$.
While in absence of the e.m. background $Q$ fluctuates around
zero, as expected by CP-invariance, fluctuations are shifted 
towards positive values as $\eisb \neq 0$. This is clearer from
Fig.~\ref{fig:1}, where we plot the corresponding distributions
of $Q$; in this case we obtain
${\avq (\vec E_I, \vec B)}/{\avqs_0} = 1.46(14)$.

\begin{figure}[t!]
\includegraphics*[width=0.49\textwidth]{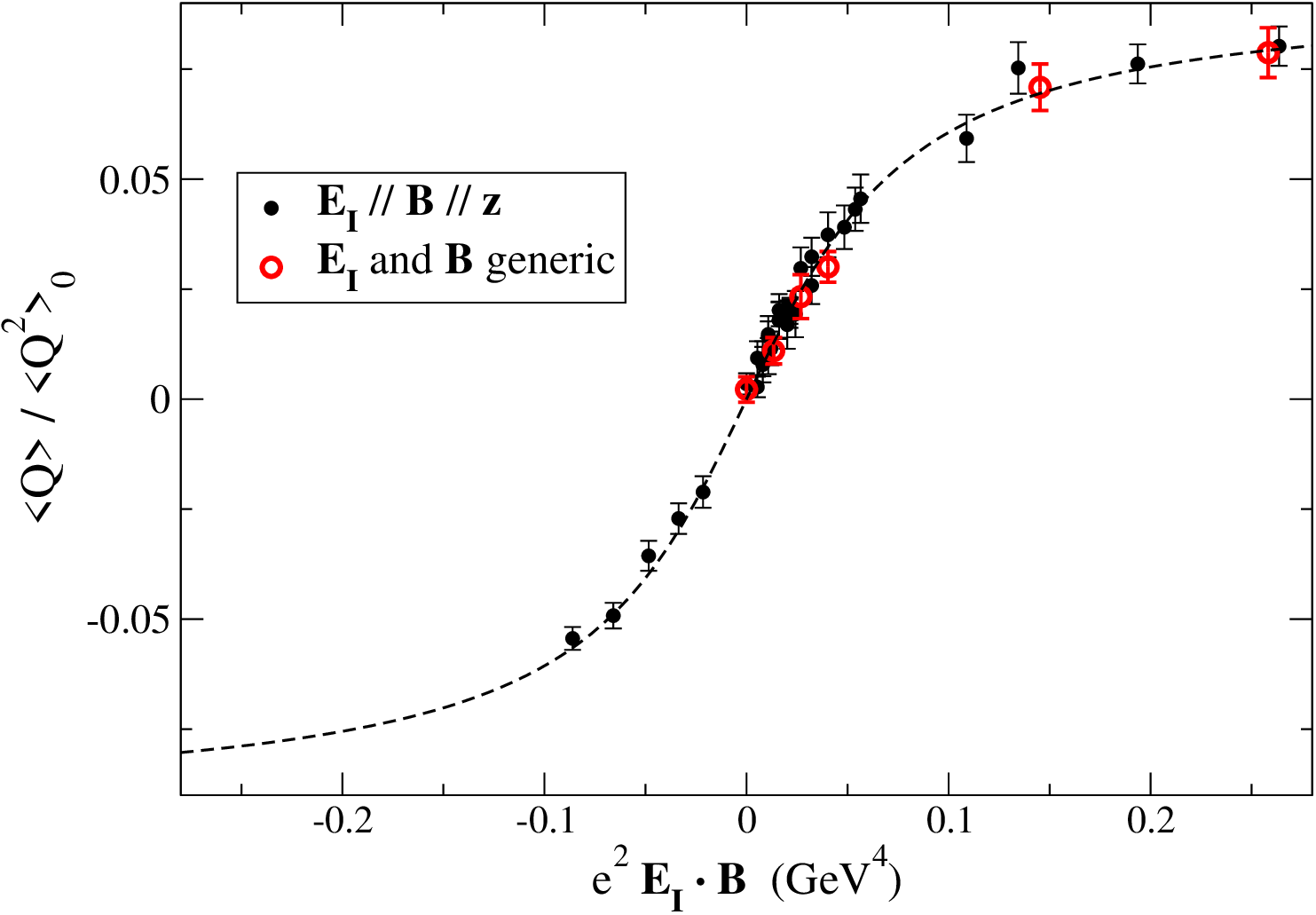}
\caption{${\avq (\vec E_I, \vec B)}/{\avqs_0}$
for various ($\vec E_I,\,
\vec B$) on a $16^4$ lattice,
$m_\pi \simeq 480$ MeV and $a \simeq 0.28$ fm.
Open circles corresponds to
$\eisb/f^2$:
$(2, -8, -2)\cdot (8, 1, 4)$,
$(1,  2,  3)\cdot (3, 2, 1)$, 
$(1,  3,  4)\cdot (2, 2, 3)$,
$(5,  5,  5)\cdot (1, 2, 3)$, 
$(6,  6,  6)\cdot (6, 6, 6)$, 
and 
$(8,  8,  8)\cdot (8, 8, 8)$. The dashed line is a best fit
to Eq.~(\ref{atanfun}).
} 
\label{fig:2}
\end{figure}

In order to 
better investigate the dependence 
of $\avq (\vec E_I, \vec B)$ on the background field values,
in Fig.~\ref{fig:2} we show
${\avq (\vec E_I, \vec B)}/{\avqs_0}$ for 
$a \simeq 0.282$ fm and for
various combinations of $\vec E_I$ and 
$\vec B$, mostly taken parallel to the $z$ axis.
All data, when plotted versus $\eisb$, 
fall on the same curve,
thus demonstrating
that $\tief$ is, 
within errors, a function of $\eisb$ alone, as expected.
We have taken different combinations
of the fields, even with $\vec E_I$ and $\vec B$
not parallel to each other, 
 having exactly the same or opposite values for $\eisb$,
thus checking also that $\tief$ is odd in $\eisb$. The dependence
is linear in $\eisb$ for small fields, then saturating
for larger fields, as common to many
systems with a linear response to external stimulation.
We have found that all data can be nicely fitted by a function
\beq
{\avq (\vec E_I, \vec B)}/{\avqs_0} = 
a_0\ {\rm atan} (a_1 \esb ) \, ;
\label{atanfun}
\eeq
the best fit curve, corresponding to $\chi^2/{\rm d.o.f.} = 0.74$, is shown
in Fig.~\ref{fig:2}.

\begin{figure}[t!]
\includegraphics*[width=0.49\textwidth]{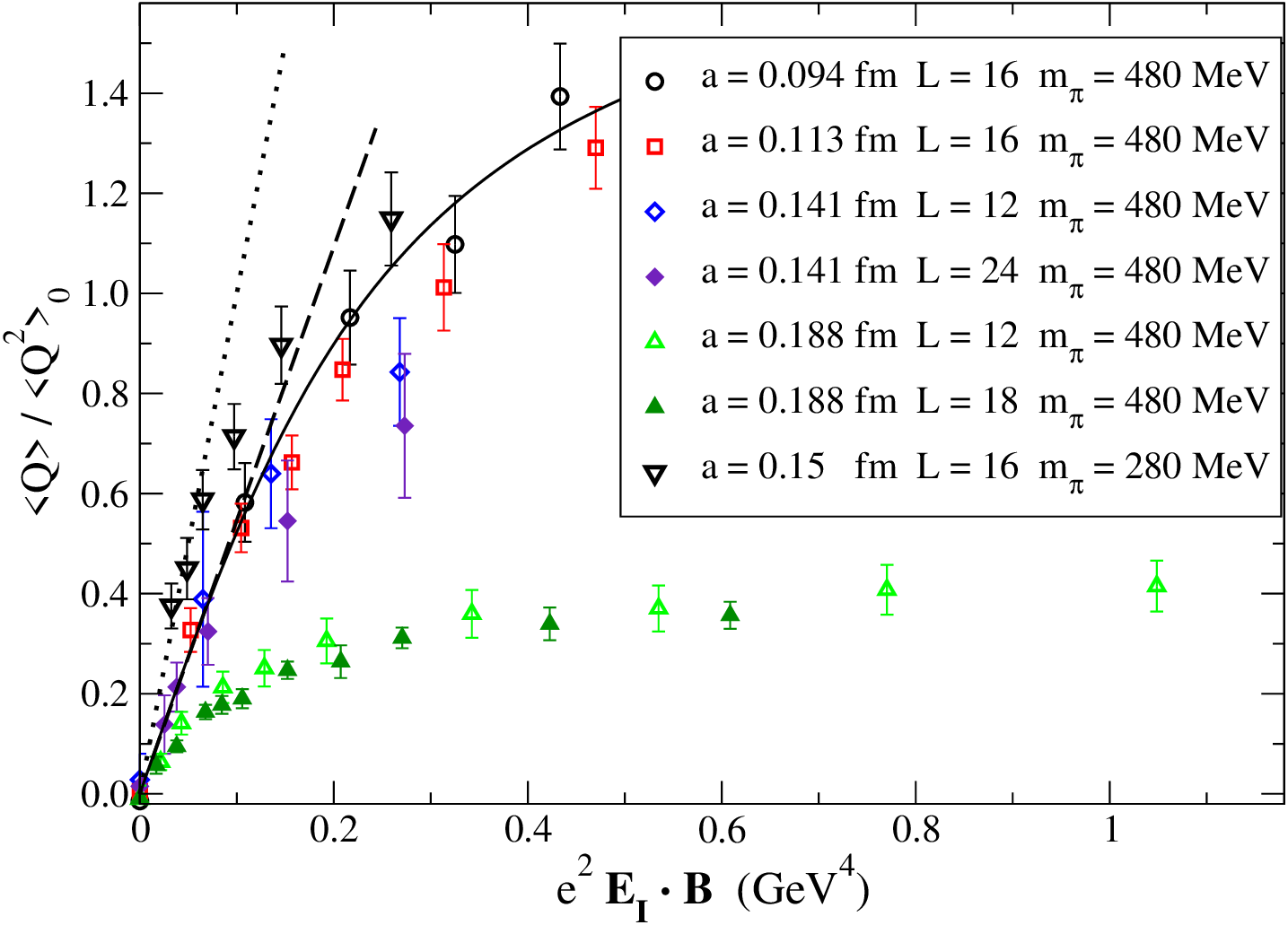}
\caption{
${\avq (\vec E_I, \vec B)}/{\avqs_0}$
for 
different lattice parameters.
The continuous and dashed lines are a best fit
to Eq.~(\ref{atanfun}) on the finest lattice and its
slope at $\eisb = 0$, for $m_\pi \simeq 480$ MeV.
The dotted line is the preliminary slope at $m_\pi \simeq 280$ MeV.
} 
\label{fig:3}
\end{figure}

To discuss finite size and UV cutoff effects,
in Fig.~\ref{fig:3} we show 
${\avq (\vec E_I, \vec B)}/{\avqs_0}$ 
for
different
spacings $a$ and lattice volumes $L^4$.
$\avq$ and $\avqs_0$ are both derivatives of the free energy with respect
to $\theta$, hence they are proportional to $V$ and,
apart from possible systematic effects, their ratio 
should be volume independent.
From Fig.~\ref{fig:3} we infer that finite size effects are 
not significant, even on the smallest volumes corresponding
to $a m_\pi L \sim 4$.

The dependence on the UV cutoff instead seems significant until
$a < 0.15$ fm. Significant lattice artifacts 
may be related to the determination of $Q$: 
if $a$
is so coarse that part of the topological background, created
by the influence of the e.m. field, lives close to the UV scale, 
then cooling may destroy part of such background.
However data obtained for $a < 0.15$ fm are in 
good agreement with each other, especially in the small
field region, which is the one relevant for the determination
of $\ceb$.

In order to determine $\ceb$, we have performed best fits of the data
in Fig.~\ref{fig:3} to the function in Eq.~(\ref{atanfun}), in a range
$e^2 \eisb < 0.8$ GeV$^4$, then
considering its slope at $\eisb = 0$ and exploiting Eqs.~(\ref{deftef})
and (\ref{deftief}). Results are reported in Table I.
We have verified that each slope
is consistent with a direct linear fit, restricted
to a narrow enough region of small $\eisb$. 
An extrapolation to the continuum limit, assuming
$O(a^2)$ corrections, yields
$\ceb = 6.09(31)$ GeV$^{-4}$ ($\chi^2/{\rm d.o.f.} \simeq 2.5$),
when including all data, 
and
$\ceb = 5.47(78)$ GeV$^{-4}$ ($\chi^2/{\rm d.o.f.} \simeq 0.1$)
if restricted to $a < 0.15$ fm. Due to the large artifacts
affecting the coarsest lattices, we prefer to quote last value
as our estimate.

Our investigation is still far from the physical
region and the chiral limit,
preliminary results obtained for 
$\beta = 5.504$ and $a m = 0.00381$ ($a \sim 0.15$ fm, $m_\pi \sim 280$ MeV)
indicate 
$\ceb \sim 10$ GeV$^{-4}$ (see Fig.~\ref{fig:3}),
suggesting that $\ceb$ increases when 
decreasing quark masses.

{\it Discussion } -- 
The phenomenological estimate
given in Ref.~\cite{mueller}
for $\kappa = 2\, \ceb$, which
is based on the effective couplings of the $\eta$ and $\eta'$ mesons 
to two photons
and to two gluons, is $\ceb \approx 0.73/(\pi^2 f_\eta^2 m_{\eta'}^2) \sim
3$ GeV$^{-4}$: 
our estimate is slightly larger, but one should consider the different systematics,
including the unphysical quark mass spectrum in our case;
moreover, it is realistic to estimate a 5\% systematic uncertainty in our knowledge
of $a$, leading an additional $\sim$ 20\% uncertainty on $\ceb$.
To give an idea of the magnitude of the effect,
we estimate  $\tef \sim 10^{-5}$
for parallel $\vec E$ and $\vec B$ with
$e B \sim 1$ GeV$^2$ and $e E \sim 1$ MeV$^2$.

Regarding the validity of analytic continuation from imaginary to 
real electric fields, we notice that, while a smooth behavior is 
expected as
the imaginary electric field approaches  zero,
at least for non-zero quark masses, 
a uniform and constant real electric field, however small,
is instead expected to induce vacuum instabilities. On the other hand, 
this is not true if we consider 
electric fields which are limited in space. 
Therefore our result should be applicable to determine the local
effective $\theta$ parameter produced by 
smooth enough but spatially limited CP-violating
e.m. fields. It would be interesting
in the future to consider the case 
of smoothly varying fields explicitly.

Finally, apart from repeating our determination with more physical
quark masses and closer to the continuum limit, 
it would be interesting to extend our investigation also to 
finite temperature, especially across and 
right above the deconfinement transition, where the determination
of the effective pseudoscalar QED-QCD interaction would be relevant
also to the phenomenology of the CME in heavy ion
collisions.

\noindent {\bf Acknowledgements:}
We thank A.~Alexandru, C.~Bonati, A.~Di~Giacomo and E.~Vicari 
for useful discussions. 
We are
grateful to Guido Cossu for his help in the determination
of the zero modes of the overlap operator.
Numerical computations have been performed on computer facilities
provided by INFN, in particular on two GPU farms in Pisa and Genoa
and on the QUONG GPU cluster in Rome.
We thank the Galileo Galilei Institute for Theoretical Physics
for the hospitality during the completion of this work.


\begin{thebibliography}{9}
\bibitem{taste}
This corresponds to the lightest pion; taste
violations are of the order of 40\% at the 
finer spacing adopted~\cite{blum}.
\bibitem{footnote1}
The Sommer parameter, measured 
for $a = 0.094$ fm on a $24^4$ lattice, is
$r_0/a = 5.36(10)$ and $r_0/a = 5.07(12)$
for zero e.m. field and $\eisb \simeq 0.12$ GeV$^4$
respectively.
\end{thebibliography}
\end{document}